\documentclass[submission,copyright,creativecommons]{eptcs}
\providecommand{\event}{PLACES 2026} 

\usepackage{iftex}
\usepackage{amsmath}
\usepackage{amssymb}
\usepackage{amsthm}
\usepackage{tikz}
\usepackage[title]{appendix}
\usepackage[UKenglish]{babel}
\usepackage{caption}
\usepackage{cleveref}
\usepackage{csquotes}

\ifpdf
  \usepackage{underscore}
  \usepackage[T1]{fontenc}
\else
  \usepackage{breakurl}
\fi

\theoremstyle{definition}
\newtheorem{theorem}{Theorem}[section]

\newtheorem{definition}[theorem]{Definition}
\newtheorem{ruleblock}[theorem]{Rule}

\usetikzlibrary{arrows.meta}
\usetikzlibrary{arrows, positioning} 
\usetikzlibrary{calc, positioning}
\usetikzlibrary{shapes.geometric}

\newcommand{\monitorTr}{\xrightarrow{m,v}}

\newcommand{\tuple}[1]{(#1)}
\newcommand{\var}[1]{\ensuremath{#1}}
\newcommand{\lists}[1]{\ensuremath{#1}}
\newcommand{\const}[1]{\textbf{\upshape{#1}}}
\newcommand{\set}[1]{\ensuremath{\mathrm{#1}}}
\newcommand{\emptyratio}{\ensuremath{\varepsilon}}

\newcommand{\code}[1]{\ensuremath{\mathtt{#1}}}

\newcommand{\returnType}[1]{\ensuremath{\mathrm{returnType}(#1)}}
\newcommand{\enumLabelsT}[1]{\ensuremath{\mathrm{enumLabels_T}(#1)}}

\newcommand{\preds}[2]{\ensuremath{\mathrm{Preds}(#1)(#2)}}
\newcommand{\update}[2]{\ensuremath{\mathrm{update}(#1, #2)}}
\newcommand{\eval}[2]{\ensuremath{\mathrm{eval}(#1,#2)}}

\newcommand{\defT}[1]{\ensuremath{\mathrm{def_T}(#1)}}
\newcommand{\actionsT}[1]{\ensuremath{\mathrm{actions_T}(#1)}}
\newcommand{\ratiosT}[1]{\ensuremath{\mathrm{ratios_T}(#1)}}
\newcommand{\attrsT}[2]{\ensuremath{\mathrm{attrs_T}(#1,#2)}}
\newcommand{\ratioT}[2]{\ensuremath{\mathrm{ratio_T}(#1,#2)}}
\newcommand{\destT}[2]{\ensuremath{\mathrm{dest_T}(#1,#2)}}
\newcommand{\preAssignT}[2]{\ensuremath{\mathrm{preAssign_T}(#1,#2)}}
\newcommand{\postAssignT}[2]{\ensuremath{\mathrm{postAssign_T}(#1,#2)}}
\newcommand{\predT}[2]{\ensuremath{\mathrm{pred_T}(#1, #2)}}
\newcommand{\decisionsT}[2]{\ensuremath{\mathrm{decisions_T}(#1, #2)}}

\newcommand{\typestate}{\ensuremath{\mathrm{T}}}
\newcommand{\send}[1]{\ensuremath{\langle#1\rangle}}
\newcommand{\receive}[1]{\{#1\}}
\newcommand{\trsT}{\ensuremath{\mathrm{Trs(T)}}}

\newcommand{\actname}[1]{\texttt{#1}}

\title{Modelling Distributed Applications with Mixed-Choice Stateful
Typestates \footnote{Work partially supported by NOVA LINCS (Grant no. UID/04516/2025) with the financial support of FCT.IP and EU Horizon Europe project TaRDIS (Grant Agreement no. 101093006).}}
\author{Francisco Parrinha \qquad João Mota \qquad António Ravara 
    \institute{NOVA LINCS and NOVA FCT, Lisbon, Portugal}
}

\def\titlerunning{Modelling Distributed Applications with Mixed-Choice Stateful Typestates}
\def\authorrunning{F. Parrinha, A. Ravara \& J. Mota}
\begin{document}
    \maketitle
\vspace{-1em}
\begin{abstract}
    \label{sec:abstract}
    \textbf{Abstract.} Distributed systems have become increasingly prevalent in the software industry. Due to their intrinsic complexity, much research has focused on the verification of their behaviour. An active research line is around behaviour models that capture these protocols -- e.g., \textit{session types}, or \textit{typestates} -- allowing their static verification. 

    Correctly designing distributed protocols is not trivial. Their communication behaviour is typically implicitly defined via asynchronous message handlers, making errors harder to detect until execution. While \textit{typestates} can ease the design process by explicitly defining correct sequences of operations, they struggle in two ways: they lack the expressiveness to define quantitative constraints that govern distributed protocols (i.e., number of acknowledgements for a quorum); and they assume strict sequencing of operations, failing to capture concurrent input/output actions in a state, typical of the distributed setting. Furthermore, runtime network failures cannot be statically verified.

    We present a probabilistic runtime solution extending \textit{typestates} with: \textit{(i)} an internal mutable state for the expression of quantitative constraints; \textit{(ii)} \textit{mixed sessions} to represent concurrent input and output actions; \textit{(iii)} expected \textit{ratios} for the number of actions in a state, with monitoring semantics to detect deviations from an expected behaviour at runtime.
    
    We demonstrate the suitability of our solution with two examples that motivated our approach: an acknowledgement protocol with a participant that sends several messages while waiting for a response, effectively modelling input and output operations in a state; and a voting protocol whose participants try to achieve consensus on a single bit using a quorum, thus, requiring an internal mutable state, while respecting a pre-defined distribution for the volume of exchanged messages.
    
    \vspace{0.5em}
\end{abstract}
\vspace{-1.7em}
\section{Introduction}
\label{sec:introduction}
Distributed systems are intrinsically complex. As their popularity has increased, so has the need for research on verifying their correctness~\cite{DBLP:conf/ecoop/HuYH08}. An essential aspect is the description of their communication protocols, being one approach based on \textit{behaviour models}~\cite{DBLP:journals/csur/HuttelLVCCDMPRT16}. Typestates~\cite{DBLP:journals/tse/StromY86} are an example. They not only capture these protocols, but also allow for the static verification of their correctness~\cite{DBLP:journals/darts/TiroreBC25}. However, traditional typestates face two key limitations for distributed systems: firstly, they lack the expressiveness to correctly define stateful operations (i.e., counting acknowledgements before transitioning); secondly, by assuming a strict sequencing of operations, they cannot model concurrent \textbf{input} and \textbf{output} operations in a single state. Furthermore, distributed systems usually rely on asynchronous message handlers for their communication on a network that may not always be reliable (e.g., channels may randomly close, partitions may occur without prior knowledge), hampering their static verification. For these situations, monitors are usually employed to observe executions and detect deviations from an expected behaviour.

Our work aims to address the presented limitations in three ways: \textit{(i)} to represent quantitative restraints in stateful operations (e.g., express a number of retries left, or required acknowledgements to trigger a transition); \textit{(ii)} with \textit{mixed sessions}~\cite{DBLP:journals/tcs/CasalMV22}, the description of concurrent behaviour of \textbf{input} (i.e., receiving messages) and \textbf{output} actions (i.e., sending messages) in a single state; \textit{(iii)} inspired by \textit{probabilistic session types}~\cite{DBLP:conf/forte/BurloFS20,DBLP:conf/coordination/BurloFSTT21,DBLP:journals/pacmpl/DasWH23}, an extension to typestate actions with probability distributions (which we call \textit{ratios}) corresponding to the expected volume of executions that each will take in the state they are in. Our objective is to provide a framework that describes protocols with stateful operations and that captures concurrent behaviour per state, while at the same time, supporting the generation of monitors that check whether a given protocol is behaving as expected at runtime. By assigning \textit{ratios} to each action in a state, the generated monitors can use a \textit{frequentist} approach to infer the runtime correctness for a given protocol~\cite{DBLP:conf/forte/BurloFS20,DBLP:conf/coordination/BurloFSTT21}, detecting runtime failures that cannot be statically verified (e.g., network failures).

Our main contribution is an extension of JaTyC typestates~\cite{DBLP:journals/scp/BacchianiBGMR22,DBLP:conf/coordination/MotaGR21} with an internal mutable state that supports the runtime data that directly governs the execution flow of distributed systems. With this, the typestates can capture stateful operations. Other key contributions are:
\begin{itemize}
    \item We rigorously define the syntax and operational semantics for our typestates to make sure transitions are respected according to the new internal state and the introduction of \textit{mixed sessions}.
    \item We extend JaTyC typestates with \textit{ratios} for the volume of action executions per state and define monitoring semantics to detect deviations from the intended protocol behaviour at runtime, providing a foundation for the generation of monitors.
    \item We demonstrate two examples with popular mechanics in distributed systems (e.g., acknowledgment and consensus), to illustrate the expressiveness of our framework.
\end{itemize}

\vspace{-1em}
\section{Motivating Examples}
\label{sec:motivating-examples}
Consider a very simple protocol: the \textit{AlternatingBitProtocol}. It contains two participants: the \textbf{sender} and the \textbf{receiver}. The \textbf{sender} stores a single bit in its internal state and repeatedly sends a message \actname{msg} with its bit, while waiting for a response from the \textbf{receiver}, an \actname{ack} message. Receiving this acknowledgment moves the \textbf{sender} to state~\var{S_1}. Here, it flips the stored bit (an internal action) and, by sending a new message \actname{msg} with a new bit value, it goes back to state \var{S_0}. Regarding the \textbf{receiver}, it always stores the last acknowledged bit in its internal state and waits for a \actname{msg} message from the \textbf{sender}, moving to state~\var{R_1}. Upon this, it replies, \actname{ack}nowledging the received bit, moving again to state~\var{R_1}.

\vspace{-0.5em}
\begin{figure}[ht]
    \[
    \begin{array}{cc}
        \textbf{Sender Graph} & \textbf{Receiver Graph} \\
        \begin{tikzpicture}[
            state/.style={
                circle,            
                draw,              
                minimum size=1cm,
                align=center,
                font=\small
            },
            ->,                  
            >=Stealth,           
            node distance=1cm 
        ]	
            \node[state] (S0) {\var{S_0}};
            \node[state, right=of S0] (S1) {\var{S_1}};
            
            \draw[->] ([xshift=-1cm]S0.west) -- (S0);
            \draw[->] (S0) edge[bend left] node[above]{$\overline{\actname{msg}}$} (S1);

            \draw[->] (S1) edge[bend left] node[below]{\actname{ack}} (S0);
            \draw[->] (S1) edge[loop above] node{$\overline{\actname{msg}}$} ();
        \end{tikzpicture}
        &

        \raisebox{0.4cm}{
            \begin{tikzpicture}[
                state/.style={
                    circle,            
                    draw,      
                    minimum size=1cm,
                    align=center,
                    font=\small
                },
                ->,                 
                >=Stealth,           
                node distance=1cm 
            ]   
                \node[state] (R0) {\var{R_0}};
                \node[state, right=of R0] (R1) {\var{R_1}};

                \draw[->] ([xshift=-1cm]R0.west) -- (R0);
                \draw[->] (R0) edge[bend left] node[above]{\actname{msg}} (R1);

                \draw[->] (R1) edge[loop right] node[right]{$\overline{\actname{ack}}$} (R1);
                \draw[->] (R1) edge[loop above] node[above]{\actname{msg}} (R1);
            \end{tikzpicture}
        }
    \end{array}
    \]
    \captionsetup{skip=-5pt}
    \caption{Graphical typestates for the AlternatingBitProtocol.}
    \label{fig:g-ts-abp}
\end{figure}
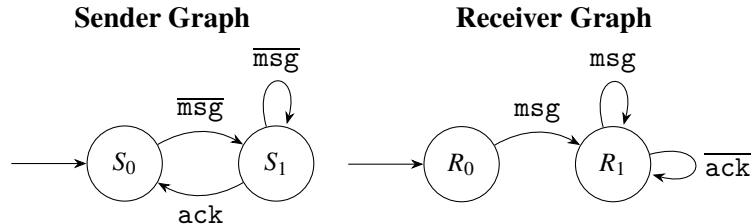
The \textit{AlternatingBitProtocol} is an acknowledgment protocol. The sender notices desynchronisation when it receives an \actname{ack} message with a bit different than the one it is sending. The receiver also knows there is an error when it receives a \actname{msg} message with a bit equal to the last acknowledged one. Both participants ignore outdated messages. The state-machines for both participants are illustrated in Figure~\ref{fig:g-ts-abp}, where labels with an overline denote \textbf{output} actions and simple labels \textbf{input} actions. 

A textual representation of the automata above needs to capture two aspects: the states and the actions available in each one. The rigorous definition of the grammar is presented later in Section~\ref{sec:grammar}, but nonetheless, we construct now the textual typestates for the \textit{AlternatingBitProtocol}. We use \enquote{$+$} to denote choice, angled brackets for output actions, and curly brackets for input actions. Therefore, our textual typestate for Figure~\ref{fig:g-ts-abp} looks as illustrated in Figure~\ref{fig:t-ts-abp}.

\begin{figure}[ht]
    \[
    \begin{array}{cc}
        \textbf{Sender Typestate} & \textbf{Receiver Typestate} \\ [0.5em]
        \begin{aligned}
            S_0 &= \send{\actname{msg}[\emptyratio]:S_1}  \\
            S_1 &= \send{\actname{msg}[\emptyratio]:S_1} + \receive{\actname{ack}[\emptyratio]:S_0}
        \end{aligned}
        &
        \begin{aligned}
            R_0 &= \receive{\actname{msg}[\emptyratio]: R_1} \\
             R_1 &= \send{\actname{ack}[0.5]:R_1} + \receive{\actname{msg}[0.5]:R_1}
        \end{aligned}
    \end{array}
    \]
    \captionsetup{skip=0pt}
    \caption{Textual typestates for the AlternatingBitProtocol \protect\footnotemark.}
    \label{fig:t-ts-abp}
\end{figure}
\footnotetext{To simplify notation, we do not use overlines in textual typestates.} 

\vspace{-1em}

We are using mixed sessions~\cite{DBLP:journals/tcs/CasalMV22} to describe the concurrent behaviour of sending and receiving messages in states \var{S_1} and \var{R_1}. We also include ratios (with \enquote{\emptyratio} denoting an inexistent ratio) to each action in state \var{R_1} to specify responsiveness requirements for the \textbf{receiver}. Ideally, the sender should receive an \actname{ack} message for each bit emission (which may have required sending several \actname{msg} messages with that bit before the receiver was able to acknowledge). However, for the receiver to be perfectly responsive, it should acknowledge one message per bit emission. So, we express that the volumes of \code{ack} (output) and \code{msg} (input) actions within the total number of action executions in the state \var{R_1} are approximately equal (e.g., fifty percent each). 
\medskip

Now consider the \textit{BitVoteProtocol}, a more sophisticated protocol. The protocol contains \var{n} \textbf{peers} and one \textbf{leader} \var{l}. The \textbf{leader} can periodically send vote requests with the message \actname{vreq}. Each vote belongs to a particular voting round. While the \textbf{leader} sends \actname{vreq} messages, it waits for an acknowledgement from the \textbf{peers}, who reply with their chosen bit with a \actname{vack} message. The \textbf{leader} may insist on the same vote round to collect all votes from the \textbf{peers}, stopping after \var{k} retries, or until all acknowledgements are collected. At this point, the \textbf{leader} broadcasts a write-back message \actname{vwb} with the most frequent bit among all received votes, starting the next voting procedure. A voting round is successful if a majority quorum is achieved. Figure \ref{fig:g-ts-bvp} shows a graphical representation for both typestates:

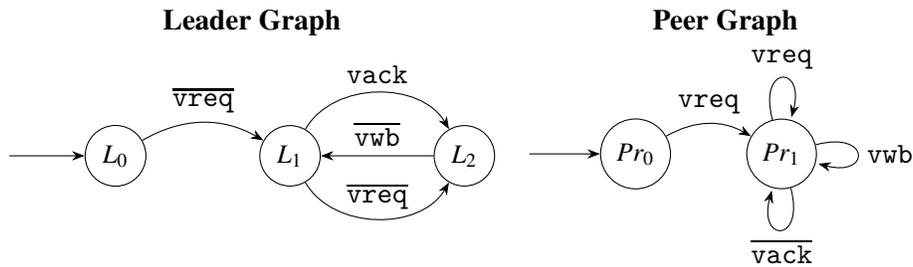
\begin{figure}[ht]
    \[
    \begin{array}{cc}
        \textbf{Leader Graph} & \textbf{Peer Graph} \\
        \raisebox{0.55cm}{
            \begin{tikzpicture}[
                    state/.style={
                        circle,
                        draw,
                        minimum size=0.5cm,
                        align=center,
                        font=\small
                    },
                    ->,
                    >=Stealth,
                    node distance=1.5cm
                ]
                \node[state] (L0) {\var{L_0}};
                \node[state, right=of L0] (L1) {\var{L_1}};
                \node[state, right=of L1] (L2) {\var{L_2}};

                \draw[->] ([xshift=-1cm]L0.west) -- (L0);
                \draw[->] (L0) edge[bend left] node[above]{$\overline{\actname{vreq}}$} (L1);

                \draw[->] (L1) edge[bend right=60] node[above]{$\overline{\actname{vreq}}$} (L2);
                \draw[->] (L1) edge[bend left=60] node[above]{\actname{vack}} (L2);

                \draw[->] (L2) edge node[above]{$\overline{\actname{vwb}}$} (L1);
            \end{tikzpicture}
        }
        &
        \begin{tikzpicture}[
            state/.style={
                circle,
                draw,
                minimum size=0.5cm,
                align=center,
                font=\small
            },
            ->,
            >=Stealth,
            node distance=1cm
        ]
            \node[state] (P0) {\var{Pr_0}};
            \node[state, right=of P0] (P1) {\var{Pr_1}};

            \draw[->] ([xshift=-1cm]L0.west) -- (P0);
            \draw[->] (P0) edge[bend left] node[above]{\actname{vreq}} (P1);

            \draw[->] (P1) edge[loop above] node{\actname{vreq}} ();
            \draw[->] (P1) edge[loop right] node{\actname{vwb}} ();
            \draw[->] (P1) edge[loop below] node{$\overline{\actname{vack}}$} ();
        \end{tikzpicture}
    \end{array}
    \]
    \captionsetup{skip=-10pt}
    \caption{Graphical typestates for the BitVoteProtocol.}
    \label{fig:g-ts-bvp}
\end{figure}

\vspace{-1em}
A textual representation for the \textbf{leader's} automata must capture three things: \textit{(i)} state \var{L_1} must produce both \textbf{input} (receiving \actname{vack}) and \textbf{output} (sending \actname{vreq}) actions; \textit{(ii)} ideally, the \textbf{leader} should receive an acknowledgement (\actname{vack}) for every vote request (\actname{vreq}) sent; \textit{(iii)} the typestate can only move from state \var{L_1} to state \var{L_2} after it received all acknowledgements or after all retries have been spent. In Figure~\ref{fig:leader-typestate} we use mixed sessions to capture the concurrent behaviour in \textit{(i)} and ratios to specify the correct volume of \actname{vreq} and \actname{vack} messages in point \textit{(ii)}. However, point \textit{(iii)} requires stateful operations. Therefore, we extend our typestates with an internal state with counters, assignment rules to update the counters, and predicates to trigger transitions using these counters. The \textbf{leader's} internal state looks as follows:
\begin{figure}[ht]
    \[
    \begin{array}{cc}
        \begin{aligned}
            \Sigma_c &= \{ n = 2,\; k = 5 \} \\
            \Sigma &= \{ acks = 0,\; retries = k \} \\
            \text{Preds} &= \{ P_1 : acks = n,\ P_2 : retries = 0 \}         
        \end{aligned}
        &
        \text{Assigns} = \left\{
                \begin{aligned}
                    & A_1 : acks \gets acks + 1, \\
                    & A_2 : retries \gets retries - 1, \\
                    & A_3 : acks \gets 0, \\
                    & A_4 : retries \gets k
                \end{aligned}
            \right\}
    \end{array}
    \]
    \captionsetup{skip=3pt}
    \caption{The internal state for the Leader.}
    \label{fig:leader-internal-state}
\end{figure}

We use $\Sigma$ and $\Sigma_c$ to define the counters a typestate may hold, with $\Sigma_c$ being used for read-only values. In Figure~\ref{fig:leader-internal-state} we define the variables \textit{acks} and \textit{retries} to denote the current number of received acknowledgements and the number of retries left; \var{n} and \var{k}, as read-only, to represent the number of peers and the starting number of retries, respectively. The map \enquote{Preds} stores predicates. These are assigned to actions in a typestate. Each predicate is evaluated whenever the associated action is executed, triggering the corresponding transition when their evaluation yields \code{true}. The map \enquote{Assigns} stores assignments that only change the values stored in $\Sigma$. We classify assignments in two categories: \enquote{pre-assignments}, which are applied to the internal state before the predicates are evaluated; and \enquote{post-assignments} that are only executed after a transition occurs.

\begin{figure}[ht]
    \[
    \begin{aligned}
        L_0 & = \send{\actname{vreq}[\emptyratio;[A_2];[]]:L_1 []} \\
        L_1 & = \send{\actname{vreq}[0.5;[A_2];[P_2]]:L_2[A_3, A_4]} + \receive{\actname{vack}[0.5; [A_1]; [P_1]]:L_2[A_3, A_4]} \\
        L_2 & = \send{\actname{vwb}[\emptyratio;[];[]]:L_1[]}
    \end{aligned}
    \]
    \captionsetup{skip=5pt}
    \caption{Textual typestate for the Leader.}
    \label{fig:leader-typestate}
\end{figure}

Each action contains now three parameters: a ratio, a list with keys of \enquote{Assigns} representing \enquote{pre-assignments} and a list with keys belonging to \enquote{Preds} for each predicate. We also add a list with \enquote{post-assignments} after the destinations. For example, the predicates~\var{P_1} and~\var{P_2} check whether all acknowledgements have been received or all retries have been spent. Either condition triggers a transition to state~\var{L_2}. The assignments \var{A_1} and \var{A_2} are \enquote{pre-assignments} and they update the number of acknowledgements and of retries, respectively, while~\var{A_3} and \var{A_4} are \enquote{post-assignments} to clear the internal state at the end of a voting round. To remind what the order of execution is: firstly, \enquote{pre-assignments} are applied, then the predicates are evaluated to trigger transitions, and if so, the \enquote{post-assignments} are executed. So, for example, in state \var{L_1}, whenever the \textbf{leader} sends a \actname{vreq} message, it firstly applies the \enquote{pre-assignment} \var{A_2}, decreasing the number of retries, then, the predicate \var{P_2} is evaluated. If the \textbf{leader} never receives a \actname{vack} message, it remains in this state until \var{P_2} yields \code{true}. Here, the \textbf{leader} moves to state \var{L_2} and applies the \enquote{post-assignments} \var{A_3} and \var{A_4}. We now correctly model stateful operations. The ratios used in state~\var{L_1} declare that there should be an approximately equal number of \actname{vreq} and \actname{vack} messages, capturing point~\textit{(ii)}.

\medskip

The typestate for a \textbf{peer} is much simpler. We only use mixed sessions in state \var{Pr_1} to denote the concurrent behaviour of sending acknowledgements while receiving vote requests and write-backs, and ratios to specify the correct volume of \actname{vack} and \actname{vreq} messages. Ideally, each \textbf{peer} sends an acknowledgement for a single request. We do not define the internal state because its contents are empty.
\begin{figure}[ht]
    \[
    \begin{aligned}
        Pr_0 & = \receive{\actname{vreq}[\emptyratio;[];[]]:Pr_1[]} \\
        Pr_1 & = \send{\actname{vack}[0.5;[];[]]: Pr_1[]} + \receive{\actname{vreq}[0.5;[];[]]:Pr_1[],\;\actname{vwb}[\varepsilon]:Pr_1[]}
    \end{aligned}
    \]
    \captionsetup{skip=5pt}
    \caption{Textual typestate for the Peer.}
    \label{fig:peer-typestate}
\end{figure}


\section{Syntax}
\label{sec:syntax}
In this section we present the syntax for our typestates. This includes the grammar and the construction of a transitions set for a given typestate. To avoid building badly formed typestates, and thus, badly formed transitions sets, we also include their well-formedness rules.

\subsection{Grammar}
\label{sec:grammar}
The following grammar describes our typestates. We use a wide tilde to denote a sequence of values. The meta-variable \var{m} ranges over the set \const{Actions} of action signatures, being the latter of the form $t\ \textit{m\_name}\ (\widetilde{t})$, where \var{m\_name} belongs to \const{ActionNames}. Let \var{s} range over the set of state names \textbf{States}, and let \var{r} denote the expected average with $r \in {[0,1]} \lor r = \emptyratio$. Let \const{Preds} be a global map of predicates and \lists{P} be a list containing keys belonging to \const{Preds}. Let~$\Sigma$ and~$\Sigma_c$ be global maps with integer values, where the latter is read-only, but modifiable externally. Let \const{Assigns} be a global map of assignments for variables belonging to $\Sigma$, and only $\Sigma$. We use~\lists{A} and~\lists{A'} as lists of assignment keys belonging to \const{Assigns}, where \lists{A} is used for \enquote{pre-assignments} and \lists{A'} for \enquote{post-assignments}. The keys for all defined maps represent the names for their corresponding values.

There are two kinds of states: input and output states. An input state is of the form $\widetilde{\receive{m[r; A; P]:w[A']}}$ and an output state is of the form $\widetilde{\send{m[r; A; P]:w[A']}}$. In both states, m is the action signature, \var{r} is the ratio,~\var{A} is the list with keys of \const{Assigns} for \enquote{pre-assignments} and \var{A'} the list with keys of \const{Assigns} for \enquote{post-assignments}. When using mixed sessions, both states can be combined using the symbol \enquote{$+$}. Note that each action signature \var{m} must be unique within a state. This constraint is also applied to the combined set of action signatures when using mixed sessions. Each state can respectively be defined as:
\[
\begin{aligned}
    \receive{m_1[r_1; A_1; P_1]&:w_1[A'_1], m_2[r_2; A_2; P_2]:w_2[A'_2], ..., m_n[r_n; A_n; P_n]:w_n[A'_n]} \quad\text{and,} \\
    \send{m_1[r_1; A_1; P_1]&:s_1[A'_1], m_2[r_2; A_2; P_2]:w_2[A'_2], ..., m_n[r_n; A_n; P_n]:w_n[A'_n]} \text{,} \quad 
\text{where } n\ge 0.
\end{aligned}
\]

For more expressiveness, our typestates also support \textbf{decision states}, which allow transitions to depend on the returned value of an action. For instance, consider an authentication protocol where a client attempts to login. The \actname{login} action returns an enumeration with two possible values: \code{success} or \code{failure}. Using a decision state, we can specify that if the login succeeds, the client transitions to an authenticated state, but if it fails, it remains in the unauthenticated state. A decision state may only appear after the colon and it has the form $\widetilde{<o: s>}$, representing all possible outcomes for an action, where each \var{o} is an outcome. We only consider enumeration and boolean values for each outcome. There must be at least one decision, and each output \var{o} must be unique:

\vspace{-1em}
\[
\begin{aligned}
    <o_1:s_1,\; o_2:s_2,\; ...,\; o_n:s_n>\quad\text{where }n > 0.
\end{aligned}
\]

Now we can model an example of a typestate for the simple authentication protocol:
\[
    \begin{aligned}
        \textit{Unauth} &= \receive{\actname{login}[\emptyratio;[];[]]:\ <\code{success}: \textit{Auth},\ \code{failure}: \textit{Unauth}> []} \\
        \textit{Auth} &= \receive{\actname{logoff}[\emptyratio;[];[]]:\textit{Unauth} []} 
    \end{aligned}
\]

The complete grammar can be described as follows:
\[
    \begin{array}{c@{\qquad}c@{\qquad}c}
        \begin{aligned}
        T &::= s = u \mid s = u\, T \\
        u &::= s_{in} \mid s_{out} \mid c \\
        c &::= s_{in} + s_{out} \mid s_{out} + s_{in}
        \end{aligned}

        &
        \begin{aligned}
        s_{in}  &::= \widetilde{\receive{m[l]: wA'}} \\
        s_{out} &::= \widetilde{\send{m[l]: wA'}}
        \end{aligned}

        &
        \begin{aligned}
        l &::= r; A; P \\
        w &::= s \mid \widetilde{\langle o: s\rangle}
        \end{aligned}
    \end{array}
    \]


\newpage
\subsection{Well-formedness}
\label{sec:well-formedness}
In this section, we present the well-formedness rules for our typestates and for their transitions set. 

\begin{definition}[State Name Resolution]
    \label{def:state-name-res}
    Let
    \[
    \defT{s} = 
    \begin{cases}
        x & \text{if} \; \typestate = \mathrm{T'} \cup \{s=x\}\quad\text{for some }\mathrm{T'}\text{ and }x \\
        s & \text{otherwise}
    \end{cases}
    \]
    be a function that returns the body of a state named $s \in \mathrm{States}$ belonging to the typestate \typestate.
\end{definition}
%
\noindent
Let $\mathrm{\pi_i}(x) = x_i$  be a projection function, where $1 \leq i \leq n$, and $x$ a tuple (i.e., $(x_1,x_2,...,x_n)$) with $n \in \mathbb{N}$

\begin{definition}[Attributes of an Action]
    \label{def:action-attrs}
    For each state $s \in \set{States}$ and $m \in \set{Actions}$ in a typestate \typestate:

    $\attrsT{s}{m} = (r,w_i,A,A',P)\ \text{  if }$
    \[
        \begin{array}{c@{\quad}c}
            \begin{aligned}
                &\defT{s}= \send{\dots, m[r;A;P]:w_iA',\dots} \text{ or } \\
                &\defT{s}= \receive{\dots, m[r;A;P]:w_iA',\dots} \text{ or } \\
                &\defT{s}= \send{\dots, m[r;A;P]:w_iA',\dots} + \receive{\dots} \text{ or }
                
            \end{aligned}
            &
            \begin{aligned}
                &\defT{s}= \receive{\dots, m[r;A;P]:w_iA',\dots} + \send{\dots} \text{ or } \\
                &\defT{s}= \receive{\dots} + \send{\dots, m[r;A;P]:w_iA',\dots} \text{ or } \\
                &\defT{s}= \send{\dots} + \receive{\dots, m[r;A;P]:w_iA',\dots}
            \end{aligned}
        \end{array}
    \]
    
\noindent
    For convenience, we declare the following functions:
    \[
    \begin{array}{c@{\qquad}c}
        \begin{aligned}
            \ratioT{s}{m} &= \mathrm{\pi_1}(\attrsT{s}{m}) \\
            \destT{s}{m} &= \mathrm{\pi_2}(\attrsT{s}{m})
        \end{aligned}
        &
        \begin{aligned}
            \preAssignT{s}{m} &= \mathrm{\pi_3}(\attrsT{s}{m}) \\
            \postAssignT{s}{m} &= \mathrm{\pi_4}(\attrsT{s}{m}) \\
            \predT{s}{m} &= \mathrm{\pi_5}(\attrsT{s}{m})
        \end{aligned}
    \end{array}
    \]
\end{definition}

\begin{definition}[Decisions of an Action]
    \label{def:action-decisions}
    \[ 
    \decisionsT{s}{m} = 
        \begin{cases}
            \{o_1,\;\dots,\;o_n \} & \text{if} \; \destT{s}{m} =\;<o_1:s_1,\dots, o_n:s_n> \text{ where } n > 0\\
            \{\code{none}\} & \text{otherwise}
        \end{cases}
    \]
\end{definition}

\begin{definition}[Labels of an Enumerable Value] 
    \label{def:enum-labels}
    Considering $t \in \set{Types}$, let:
    \[ 
    \enumLabelsT{t} = 
        \begin{cases}
            \{l_1,\;\dots,\;l_n\} & \text{if} \ t = \code{enum}\ C\text{, where each l is an element of } C \text{ and } n > 0 \\
            \{\code{true}, \code{false}\} & \text{if} \; t = \code{boolean}\\
            \{\code{none}\} & \text{otherwise}
        \end{cases}
    \]
\end{definition}

\begin{definition}[Actions of a State]
    \label{def:state-actions}
    For each state $s \in \mathrm{States}$ in a typestate \typestate:
    \[
    \actionsT{s} =
    \begin{cases}
        \{ m_1, \dots, m_n \}, &
        \text{if } 
        \begin{aligned}[t]
            &\defT{s}= \send{m_1[l]:w_1[A'_1],\dots,m_n[l]:w_n[A'_n]}\\
            &\text{or } \defT{s}= \receive{m_1[l]:w_1[A'_1],\dots,m_n[l]:w_n[A'_n]}
        \end{aligned} \\[0.5em]
        \begin{aligned}
            &\actionsT{s_{in}}\ \cup \\
            &\actionsT{s_{out}}
        \end{aligned}, &
        \text{if } \defT{s}= s_{\text{in}} + s_{\text{out}} \text{ or } \defT{s}= s_{\text{out}} + s_{\text{in}} \\
        \varnothing, & \text{otherwise}.
    \end{cases}
    \]
\end{definition}	

\begin{definition}[Defined Ratios of a State]
    \label{def:state-ratios}
    For each state $s \in \mathrm{States}$ in a typestate \typestate, let
    \[
    \ratiosT{s} =
    [\, \ratioT{s}{m} \mid m \in \actionsT{s}
        \land \ratioT{s}{m}\ \neq \varepsilon \,].
    \]
    be a function that returns a list with ratios belonging to a state $s$ where $r \neq \varepsilon$.
\end{definition}


\begin{definition}[Well-formed Typestate]
\label{sec:well-formedness-typestate}
A typestate \typestate{} is considered to be well-formed if the following rules hold:
    
\begin{description}
    \item[No Duplicate State Name] $\forall s,\ u,_1,\ u_2, \quad s = u_1 \in \typestate,\ s = u_2 \in \typestate,\ u_1 = u_2$
    \item[Valid Ratio Sum] $ \forall s = u \in \typestate, \quad \ratiosT{s} \neq [] \Rightarrow (\sum_{r_i \in \ratiosT{s}} r_i) = 1 $
    \item[Enumerate All Decisions]
    \(
    \begin{array}[t]{ll}
    \forall s = u \in \typestate, & m \in \actionsT{s}, \\
        & \decisionsT{s}{m} = \enumLabelsT{\returnType{m}}
    \end{array}
    \)
\end{description}
\end{definition}


\medskip

The transitions set \trsT{} contains all transition relations for a given typestate. This set is later used to build a graph representing the protocol defined by the typestate.

\begin{definition}[Transition Set]
\label{def:trsT}
Consider a typestate \typestate{}. 
\(
\trsT \subseteq \mathrm{States} \times \mathrm{Actions}  \times \mathrm{Values} \times \mathrm{States}
\)
is its transitions set, which can be built in the following way:
\[
\begin{aligned}
    \trsT &= \{(s,m, \code{none}, s') \mid \exists u. s = u \in \typestate \land m \in \actionsT{s} \land \destT{s}{m} = s' \} \ \cup \\
    &\{(s,m, o_i, s'_i) \mid \exists u. s = u \in \typestate \land m \in \actionsT{s} \land \destT{s}{m} = <o_1:s'_1,\dots,o_n:s'_n>\}
\end{aligned}
\]

where $n \geq 1$ and $1 \leq i \leq n$.
\end{definition}

Note that the first declared state in a typestate is considered to be the \code{start} state. 
Let T be a well-formed typestate. The set of transitions Trs(T) is also well-formed if the following rules hold. 

\begin{definition} [Reachability Predicate] 
    \label{fun:is-reachable}

    Let the following function be defined as
    \[
        \mathrm{isReach}(s, \typestate, V) = 
        \begin{cases}
            \code{true} &\text{if } s \text{ is \code{start}}\\
            \code{false} &\text{if } s \in V \lor V = \emptyset \\
            \bigvee_{(s',m,v,s) \in \trsT \land s' \notin V} \mathrm{isReach}(s', \typestate, V \cup \{s\}) &\text{otherwise}
        \end{cases}
    \]	

\end{definition}

\begin{definition} [Productivity Predicate]

    \label{fun:is-productive}
    \[
        \mathrm{isProd}(s, \typestate, V) = 
        \begin{cases}				
            \code{true} &\text{if } \neg \exists(s,m,v,s') \in \trsT  \\
            \code{false} &\text{if } s \in V \lor V = \emptyset \\
            \bigvee_{(s,m,v,s') \in \trsT \land s' \notin V} \mathrm{isProd}(s', \typestate, V \cup \{s\}) &\text{otherwise}
        \end{cases}
    \]	
\end{definition}

\begin{ruleblock} [Useful States]
\label{rule:useful-states}
    \[
    \forall s = u \in \typestate,\ \mathrm{isReach}(s, \typestate, \{\}) = \code{true} \land (\exists s' = u \in \typestate.\neg \exists(s',m,v,s'') \Rightarrow \mathrm{isProd}(s, \typestate, \{\}) = \code{true})
    \]
\end{ruleblock}

\begin{ruleblock} [Deterministic]
\label{rule:deterministic}
    \[
    \forall s,m,v_1,v_2,s_1,s_2, \quad (s,m,v_1,s_1) \in \trsT \land (s,m,v_2,s_2) \in \trsT \Rightarrow s_1 = s_2 \land v_1 = v_2
    \]
\end{ruleblock}


The following properties are a direct consequence of the defined rules for both typestates and their transitions set. Typestates are \textit{weakly connected}:
\[
\begin{aligned}
    \forall s, s',&\quad s \neq s' \Rightarrow \forall i \in \{0,\dots,n-1\} \ \exists s_i,m_i,v_i,s_{i+1}.s=s_0 \land s'=s_n \ \land \\
    &((s_i,m_i,v_i,s_{i+1}) \in \trsT \lor (s_{i+1},m_i,v_i,s_i) \in \trsT), \quad  \text{where } n > 0.
\end{aligned} 
\]

\noindent
Decision states defined in typestates are totally enumerated in their transitions-set:
\[
\forall s,m,v,\quad s = u \in \typestate,\ m \in \actionsT{s},\ v \in \decisionsT{s}{m},\ \exists s'. (s,m,v,s') \in \trsT
\]

\noindent
Transitions correctly represent non-enumerable:
\[
\forall (s,m,v,s') \in \trsT, m \in \actionsT{s} \wedge (\forall s'',\ \destT{s}{m} = s'' \Rightarrow v = \code{none} \land s'' = s') 
\]

\noindent
Transitions correctly represent enumerable types:
\[
\forall (s,m,v,s') \in \trsT, m \in \actionsT{s} \wedge (\forall \widetilde{o:s}, \destT{s}{m} =\;<\widetilde{o:s}> \Rightarrow v : s' \in \widetilde{o:s})
\] 

\section{Semantics}
\label{sec:semantics}
In this section we present the operational semantics for our typestates. We demonstrate how typestates evolve: how their internal state changes with assignments and how transitions are triggered using predicates. All relations are deterministic and identifiable by an action signature \var{m} and its returned value~\var{v}. 

\subsection{Triggering Transitions}
\label{sec:triggering-transitions}
Consider the following typestate with the defined internal state:
\[
    \begin{array}{cc}
        \textbf{Internal State} & \textbf{Typestate} \\[4pt]
        \begin{array}[t]{rl}
            \set{Assigns} &= \{A_1: \textit{acks} \gets \textit{acks} + 1,\ A_2: \textit{acks} \gets 0 \} \\ [3pt]
            \set{Preds} &= \{P_1: \textit{acks} = 2 \} \\ [3pt]
            \Sigma &= \{\textit{acks} = 0\}
        \end{array} 
        &
        \begin{array}[t]{rl}
            S_0 &= \receive{\actname{m}[\emptyratio;[A_1];[P_1]]:S_1[A_2]} \\ [3pt]
            S_1 &= \send{\actname{m}[\emptyratio;[];[]]:S_0[]} 
        \end{array}
    \end{array}
\]
When the action \actname{m} is executed, the pre-assignment \var{A_1} is applied, increasing the value of \textit{acks}. However, the transition to \var{S_1} should not be triggered because the predicate \var{P_1} still yields \code{false}. We express this behaviour by keeping the typestate in the same state (e.g., \var{S_0}). After a second execution of \actname{m}, the counter \textit{acks} finally reaches the correct value for \var{P_1} (e.g.,~$\textit{acks} =~2$) to yield \code{true}, moving the typestate to state \var{S_1}. Upon this, the post-assignment \var{A_2} is applied to clear the number of \textit{acks}. We specify this behaviour with two inference rules: \textit{non-triggering} transitions, and \textit{triggering} transitions, respectively. Note that these are not probabilistic. Non-triggering transitions accumulate changes in the internal state when a given predicate is not (yet) satisfied; and triggering transitions move the system to a new state when the internal state satisfies a given predicate. Although our transitions are not probabilistic, the dependence on predicates to determine choices also appears in the work of Zhou et al.~\cite{DBLP:journals/pacmpl/00020HNY20}.

\begin{definition}[Typestate Internal State] 
    \label{def:typestate-internal-state}
    Let
    \(
    \set{Vars} \triangleq \set{Names} \mapsto \mathbb{Z}
    \)
    be the type for maps from names to integer values, where $\Sigma \in \set{Vars}$.
\end{definition}
\begin{definition}[Transition Info]
    \label{def:transition-info}
    Let
    \( 
    \mathrm{TInfo} \triangleq  \mathrm{States} \times \mathcal{P}(\mathrm{Vars}) 
    \)
    be the set containing the correct information (a pair with a state name and an internal state) for parameterized transitions.
\end{definition}

\begin{definition}[Update Typestate Context]
    \label{fun:update}
    To update an actual state $\Sigma \in \set{Vars}$ using a list of assignment names, or keys of $\mathrm{Assigns}$, an \emph{update} function can be defined as:
    \[
    \begin{aligned}			
        \update{()}{\Sigma} &= \Sigma \\
        \update{(u_1,\;\dots,\;u_n)}{\Sigma} &= (\mathrm{Assigns}(u_n)\;\circ\;\dots\;\circ\;\mathrm{Assigns}(u_1))(\Sigma)\text{,} \;\;\; \text{for } n > 0 
    \end{aligned}
    \]
    where $\forall{i} \in \{1,\dots,n\},\;u_i \in \mathrm{dom}(\mathrm{Assigns})$.
\end{definition}

\begin{definition}[Predicate Evaluation]
    \label{fun:eval} 
    The following function evaluates a sequence of predicates given their labels and the current internal state for a typestate:
    \[
    \begin{aligned}
        \eval{()}{\Sigma} &= \code{true} \\
        \eval{(p_1, \ldots, p_n)}{\Sigma)} &= \bigwedge_{i=1}^{n} \preds{p_i}{\Sigma}, \quad \text{for } n > 0
    \end{aligned}
    \]
    where $\forall{i} \in \{1,\dots,n\},\;p_i \in \mathrm{dom}(\mathrm{Preds})$.
\end{definition}

\begin{definition}[Parameterized Transition Relation] 
    \label{rel:parameterized-transition}
    For a given typestate \typestate, considering an action signature $m$ and a returned value $v$, let
    \(
    \overset{m,v}{\Longrightarrow} \;\subseteq\; \mathrm{TInfo} \times \mathrm{TInfo}
    \)
    be the relation inductively defined by the following rules: 
    \[
    \textsc{Non-Triggering} \quad
    \frac{
        \tuple{s, m, v, s'} \in \trsT \quad \Sigma' = \update{\preAssignT{s}{m}}{\Sigma} \quad \neg \eval{\predT{s}{m}}{\Sigma'}
    }{
        \typestate \vdash \tuple{s,\Sigma} \overset{m,v}{\Longrightarrow}  \tuple{s, \Sigma'}
    }
    \]

    \[
    \textsc{Triggering} \quad
    \frac{
        \tuple{s, m,  v, s'} \in \trsT \quad \Sigma' = \update{\preAssignT{s}{m}}{\Sigma} \quad \eval{\predT{s}{m}}{\Sigma'}
    }{
        \typestate \vdash \tuple{s,\Sigma} \overset{m,v}{\Longrightarrow}  \tuple{s', \update{\postAssignT{s}{m}}{\Sigma'}}
    }
    \]
\end{definition}
\medskip
 
Note that there is a strict order for the assignment and predicate operations. Firstly, the internal state is updated with the pre-assignments, then the predicates are evaluated. Post-assignments are only applied in \textit{triggering} transitions, after a transition occurs. 


\subsection{Monitoring}
\label{sec:monitoring}
The following inference rule describes monitoring behaviour. For an action signature~\var{m} and a state name~\var{s}: let $\mu_{m,s}$ be the expected ratio, $\hat{\mu}_{m,s}$ be the current ratio, $E_{m,s} \in \mathbb{R}_{\ge 0}$ be a global positive real variable for the maximum acceptable error, used to create a confidence interval~\var{I_{m,s}} around~$\mu_{m,s}$, and~$p_{m,s}$ be the number of executions an action with signature~\var{m} took in a state named~\var{s}. Let~$n_s$ be the total number of action executions in a state named~$s$, and~\var{M} be the set with all monitored information.

\begin{definition}[Monitor Log] 
    \label{def:monitor-log}
    Let
    \(
    \set{Log} \triangleq \set{States} \times \set{Actions} \times \set{Interval} \times \mathbb{R}
    \)
    be the type of an individual log entry in the monitor set $M$.
\end{definition}

\begin{definition}[Monitor Transition Info] 
    \label{def:monitor-transition-info}
    Let
    \( \set{MTInfo} \triangleq  \set{States} \times \mathbb{N} \times \mathbb{N} \times \mathcal{P}\set{(Log)}\)
    be a tuple with the correct information for the execution of a monitor. 
    \enquote{MTInfo} contains a state name, the total number of action executions in that state, the number of times a specific action was executed in that state, and a set with all monitored information.
\end{definition}

\medskip
\begin{definition}[Monitor Transition Relation] 
    \label{rel:monitor}
    For a given typestate \typestate, considering an action signature $m$ and a returned value $v$, let
    \(
    \monitorTr \subseteq \set{MTInfo} \times \set{MTInfo}
    \)
    be the relation inductively defined by the following rule:
    \vspace{-0.5em}
    \[
    \textsc{Monitoring} \quad
    \frac{
        \begin{array}{c}
            (s,m,v,s') \in \trsT \quad
            \typestate \vdash \tuple{s,\Sigma} \overset{m,v}{\Longrightarrow} \tuple{s', \Sigma'} \quad
            \hat{\mu}_{m,s} = \frac{p_{m,s} + 1} {n_s + 1} \quad \\[0.5em] 
            \mathrm{ratio_T}(s,m) \neq \emptyratio \quad
            \mu_{m,s} = \mathrm{ratio_T}(s,m)  \quad I_{m,s} = [\mu_{m,s} - E_{m,s}, \, \mu_{m,s} + E_{m,s}] 
        \end{array}
    }{
        \typestate \vdash \tuple{s,n_s, p_{m,s}, M} \monitorTr \tuple{s', n_s + 1, p_{m,s} + 1, M \cup \{\tuple{s,m,I_{m,s}, \hat{\mu}_{m,s} }\}}
    }
    \]	
\end{definition}


The \textit{monitoring} rule does not trigger transitions. It relies on pre-existing \textit{parameterized transition relations}~\ref{rel:parameterized-transition} and stores the correct information to estimate behaviour correctness at runtime at each action execution.
Using the same methodology and formulas presented in the work of Christian Bartolo Burlò et al.~\cite{DBLP:conf/coordination/BurloFSTT21}, a monitor validates the calculated ratio every time the \textit{monitoring} rule is triggered, logging deviations from the expected behaviour and illegal action executions. The implementation of a monitor uses the \textit{monitoring} rule and computes confidence intervals (CIs) around the pre-defined ratios, checking at each action execution if the current ratio fits in the calculated CI. A deviation is detected when the current ratio does not fit in the CI. Note that actions with an empty ratio (\emptyratio{}) are not monitorable.




\section{Related Work}
\label{sec:related-work}
Probabilistic session types \cite{DBLP:conf/forte/BurloFS20,DBLP:conf/coordination/BurloFSTT21} present a runtime solution for the verification of binary session types~\cite{DBLP:journals/iandc/Vasconcelos12} augmented with probability distributions over choice points. Monitors observe the exchange of messages between two parties and estimate the probabilities for the corresponding choices, comparing these with pre-assigned probabilities using confidence intervals, and issuing warnings when the runtime behaviour deviates from the expected behaviour. While our work shares the probabilistic approach as seen in~\cite{DBLP:conf/coordination/BurloFSTT21}, we focus on the behaviour of individual typestates and they do not necessarily require duality. Additionally, their symbol \enquote{*} can also be used for one-sided monitoring -- to ignore warnings when the computed average is too high~(e.g., [0.5,*]), or when it is too low (e.g., [*, 0.5]) -- which is different than the sign~\enquote{\emptyratio{}} that we use. Our monitoring semantics completely ignores actions whose ratios are~\enquote{\emptyratio}.

Mixed sessions~\cite{DBLP:journals/tcs/CasalMV22} extend traditional binary session types by allowing input and output operations to coexist in the same state, rather than strictly alternating between both. This enables more flexible protocol specifications that can better model asynchronous and concurrent communication patterns. We adopt the concept in our typestates by allowing states to have both input and output operations within a given state, combining them with the sign \enquote{$+$}.

Typestates~\cite{DBLP:journals/tse/StromY86} were introduced to track the state-dependent behaviour of objects, encoding the sequences of valid operations based on the states an object may have, raising errors when their corresponding protocols are violated. Our work extends traditional typestates with mixed sessions for the description of concurrent behaviour per state, with ratios for runtime monitoring, and with an internal state containing variables, assignment rules and predicate rules for stateful operations.

JaTyC~\cite{DBLP:journals/scp/BacchianiBGMR22, DBLP:conf/ecoop/BacchianiBGMR24,DBLP:conf/coordination/MotaGR21}, a plugin for the Checker framework, is a tool that statically verifies Java source code with respect to typestates, performing compile-time analysis to ensure objects follow the protocol of their associated typestate. In contrast, our work targets runtime monitoring to address challenges that cannot be statically verified, such as network failures and message loss, which is suitable for the concurrent and asynchronous environment of distributed systems. 

\section{Conclusion and Future Work}
\label{sec.conclusion}
Our work serves as a foundation for a monitor generation system. The formalisms we provide for our typestates not only ease the process of creating protocols by abstracting their specifications away from their low-level implementation details, but also provide guarantees regarding the correctness of the generated code for the corresponding monitors. Our typestates also help detect design errors in badly constructed protocols (i.e., those that do not follow our well-formedness rules). By adding ratios to typestate actions, monitors can validate the runtime behaviour of protocols. Furthermore, by including an internal state and mixed sessions \cite{DBLP:journals/tcs/CasalMV22}, they can better model stateful operations (i.e., triggering transitions that depend on the values of certain counters) and concurrent behaviour in single states, respectively.


We are building a Java tool that defines, validates and generates monitors from our typestates. However, since our typestates include an internal state and the generated monitors do not have direct access to the source code of the protocol, inconsistencies may arise between them. For instance, a rejected message by the protocol might still update the internal state of the monitor, causing divergence. To address this, we aim to provide a library with extensions (e.g., custom exceptions) that allow developers to notify the monitors of such cases. With these features, developers can handle the desynchronization between the monitors and their corresponding protocols. The tool can then be used by system administrators, who may choose the best course of action when incorrect behaviour is detected in their distributed systems. 

 
    \newpage
    \nocite{*}
    \bibliographystyle{eptcs}
    \bibliography{generic}
    \newpage
\end{document}